\documentclass[aps,pre,amsmath,amsfonts,floatfix,superscriptaddress,nofootinbib]{revtex4}
\usepackage{mathrsfs} \usepackage{graphicx,color} \usepackage{bbm} \usepackage{multirow}

 \let\b=\beta  \let\d=\delta
\let\e=\varepsilon  \let\h=\eta 
\let\l=\lambda    
  \let\f=\varphi 
   \let\G=\Gamma
 \let\Th=\Theta  
   
\let\ee=\epsilon \let\r=\rho \let\th=\theta \let\io=\infty

\def\FF{{\cal F}}

\def\DD{{\cal D}}\def\GG{{\cal G}} \def\SS{{\cal S}}

  \def\erf{\text{erf}}
\def\ol{\overline}

\def\to{\rightarrow} \def\la{\left\langle} \def\ra{\right\rangle}

\newcommand{\beq}{\begin{equation}} \newcommand{\eeq}{\end{equation}}
\newcommand{\wh}{\widehat} 
\newcommand{\Tr}{\text{Tr}}

\begin{document}

\title{
Exact theory of dense amorphous hard spheres in high dimension \\
I. The free energy
} 

\author{Jorge Kurchan}
\affiliation{
P.M.M.H. \'Ecole Sup\'erieure de Physique et 
Chimie Industrielles, 10 Rue Vauquelin, 75231 Paris
 Cedex 05, France
}

\author{Giorgio Parisi}
\affiliation{Dipartimento di Fisica,
Sapienza Universit\'a di Roma,
INFN, Sezione di Roma I, IPFC -- CNR,
P.le A. Moro 2, I-00185 Roma, Italy
}

\author{Francesco Zamponi}
\affiliation{LPT,
\'Ecole Normale Sup\'erieure, UMR 8549 CNRS, 24 Rue Lhomond, 75005 France}

\begin{abstract}
We consider the theory of the glass transition and jamming of hard spheres in the large space dimension limit.
Previous investigations were based on the assumption that the probability distribution within a ``cage''  is Gaussian, which is not fully consistent with numerical
results. Here we perform a replica calculation { without making any assumption on the cage shape}. We show that thermodynamic functions 
turn out to be exact within the Gaussian ansatz -- provided one allows for arbitrary replica symmetry breaking ---  and indeed agree well
with numerical results. The actual structure function (the so-called non-ergodic parameter) is not Gaussian, 
an apparent paradox which we discuss. 
In this paper we focus on the free energy, future papers will present the results for the structure functions and
a detailed comparison with numerical results.
\end{abstract}

\maketitle

\section{Introduction}

Hard spheres in the limit of large spatial dimensions provide us with an opportunity for an analytic solution covering many aspects of the 
liquid and glass physics~\cite{FRW85,KW87,FP99,PS00}. The reason why this limit is solvable is geometric: consider three spheres A, B and C, with AB and BC in contact, respectively.
What are the chances that A and C will themselves also be in contact? In high dimensions, vanishingly  small. 
This led to the realization~\cite{FRW85} that all terms in the virial expansion above the second could be neglected in high dimensions, as they involve 
geometrically heavily suppressed ``coincidences'', leaving one with only the two first terms of the series for the entropy 
 \beq\label{1SS}
\SS[\r( x)] = \int d^d x \, \r( x) [ 1 - \log \r( x) ] + \frac12 
\int d^d x d^d y \,  \r(x) \r( y) f(x -  y) \ .
\eeq
Here $f(x) = e^{-v(x)} - 1 = -\theta(D - |x|)$ is the Mayer function of the Hard Sphere potential $v(x)$, which is infinite for $|x|<D$ and zero otherwise.
$D$ is therefore the sphere diameter.
For the liquid phase, one has $\r(x)=\r$, a constant, and the above expression gives the liquid entropy; non-uniform phases are described by solutions
of the stationarity equation $\partial \SS/\partial \r(x) = 0$ that are not constant\footnote{This statement is exact at low densities. At higher densities Eq.~(\ref{1SS}) would have corrections, 
however these corrections are exponentially small in the density region we consider in this paper~\cite{FP99,PS00}.}.
The liquid phase stays metastable at higher densities 
when one expects the thermodynamics to be dominated by a modulated, crystalline phase, which is however only known in small dimensions~\cite{ConwaySloane}.

Even neglecting the crystalline phase, one expects  that at  some density, there is the possibility of a thermodynamic glass transition into a phase where one or several 
spatially-dependent non periodic solutions dominate. 
The conceptual (and practical) method to neglect the crystalline phase and uncover a possible liquid-glass transition was proposed years ago~\cite{KT89,Mo95}:
one studies the system perturbed by a spatially random external field -- whose function is to kill the crystal and select one of many amorphous solutions. One in fact
computes the average over the ``pinning'' field realizations, and then continues the
solution to zero field intensity. This program has been followed for the hard sphere case~\cite{MP09} (see \cite{PZ10} for the state of the art). 
The inclusion of a random field brings about the problem of treating quenched averages, and this has been done using replica methods (i.e. introducing $m$ copies of the same system).
One ends up with a truncated virial expansion~\cite{PZ10}
\beq
\SS[\r(\bar x)] = \int d\bar x \r(\bar x) [ 1 - \log \r(\bar x) ] + \frac12 
\int d\bar x d\bar y \r(\bar x) \r(\bar y) f(\bar x - \bar y)
\label{virial}
\eeq
where now the density $\r(\bar x)$ is function of $m$ coordinates in $d$ dimensions ${\bar x}=\{x_a\}=\{x_1, ...,x_m\}$, and one has then to analytically continue over $m$ from integer values of $m$ to non-integer ones. The stationary points of Eq.~\ref{virial} satisfies the equation
\beq
\log \r(\bar x) ] = 
\int  d\bar y  \r(\bar y) f(\bar x - \bar y) \ .
\label{equation}
\eeq
This equation has also a probabilistic meaning, as discussed in~\cite{MPTZ11}. Indeed the same equation appears in the study of a hard sphere model on the Bethe lattice~\cite{MKK08}.

Even though equation (\ref{equation}) is (morally) exact in large dimensions, in order for it to be useful we need an expression for $\r$ that contains $m$ explicitly and allows to  continue the results  for real  $m$.
An approach to do this is to propose a ``Gaussian ansatz'' for  $\r(\bar x)$~\cite{MP09,PZ10}:
\beq
\r(\bar x) = \frac{\r m^{-d/2}}{(2\pi A)^{(m-1) d /2}} \exp\left({-\frac{1}{2mA} \sum_{a<b}^{1,m} (x_a - x_b)^2} \right)
\label{Gaussian}
\eeq
 and to extremize (\ref{virial}) with respect to the parameters in (\ref{Gaussian}).
 Having to content oneself with the  Gaussian ansatz might seem somewhat disappointing:  we have payed the price of going to  unphysically high dimensions in order to 
 have an exact answer, and now we do not even have this.  
 
The purpose of this paper is that of obtaining the exact solution of (\ref{equation}) in the high-dimensional limit without assumptions. We find that 
 the Gaussian ansatz turns out to be, in a sense, exact: it gives for large dimensions the exact values for the thermodynamic quantities. 
The reason, whose consequences we shall develop below, can be seen  as follows: a generic replica problem is written in terms of the order parameter $\r(\bar x)$, or, alternatively,
of the  tensors $\langle x_a x_b \rangle, $$ \langle x_a x_b x_c \rangle$, $\langle x_a x_b x_c x_d \rangle, ... $ The solution involves  an ansatz for replica tensors of all degrees, which makes
the problem analytically hard.
 However, in our case, the $x_a$ are vectors in $d$-dimensional space, and we are looking for a solution that is statistically translationally invariant and isotropic.
 The only possibility with these properties is that $\r(\bar x)$ depends exclusively on the   $  |x_a-x_b|^2 = q_{aa} + q_{bb} -2 q_{ab} $, where we have introduced the scalar products $q_{ab}=x_a \cdot x_b$.
 All $d$-dimensional integrals may be expressed as {\em low dimensional} integrals in terms of the $q_{ab}$, with volume factors,
  a simple generalization of spherical coordinates.
 As we shall see, the  dimensionality appears explicitly, and the limit of large $d$ may be taken in a straightforward way, by saddle point evaluation of the integrals.

 It will turn out, however, that  the ``cage shape'' is not Gaussian, as already observed in simulations \cite{CIPZ12}, but may be calculated exactly for large $d$ within this framework. 
 The fact that the Gaussian approximation gives the correct result for thermodynamic functions but by itself does not give the right cage shape may be understood with a simple example.
 Consider a single particle in a $d$-dimensional spherical potential $V$, which for convenience we write as  $\beta V= d \, U (|x|^2/d)$.  Let us denote $q \equiv r^2 = |x|^2$.
 Clearly, the exact Gibbs distribution is 
 \beq
 \r(q)= \frac{e^{-d \, U(q/d)}}{ {\cal{N}}_o } \;\;\; \text{with} \;\;\;  {\cal{N}}_o= \frac{1}{2} \int dq \; q^{(d-2)/2} \; e^{-d \, U(q/d)}
 \eeq
 The entropy $S = - \int dx \; \r \ln \r$ is easily evaluated and gives:
 \beq
 S = - \int dq \; q^{(d-2)/2} \;  \frac{e^{-d \, U(q/d)}}{{ {\cal{N}}_o } }  \left\{ -d \, U(q/d)  -  \ln {\cal{N}}_o \right\}
 \eeq
 For large dimension $d$, the integrals for ${ {\cal{N}}_o }$ and for $S$ are dominated by the saddle point $q^*$ which maximizes \\
 $\frac{1}{2}{\ln q} -U(q/d)$. The entropy is then, to leading order in $d$:
  \beq
 S =  -d \, U(q^*/d)  -  \ln {\cal{N}}_o \sim  -d \, U(q^*/d)  +d \, U(q^*/d) + \frac{d}{2}{\ln q^*} = \frac{d}{2}{\ln q^*} 
 \eeq
 Suppose instead that we had done this calculation approximately, proposing a variational Gaussian distribution $\rho_G = e^{-|x|^2/(2A)}/ {\cal{N}}_G$.
 The same calculation as before gives for the entropy:
  \beq
  S_G = - \int dx \; \r_G \ln \r_G  \sim \frac{d}{2}{\ln q_A} 
\eeq 
where $q_A$ maximizes  $\frac{1}{2}{\ln q} - q/(2 d \, A)$.
We now have to fix  the parameter $A$ by minimizing the free energy:
 \beq
-\b F= \int dx \;  [-\b V - \ln \r_G] \; \r_G \sim -d \, U(q_A/d)+  \frac{d}{2}{\ln q_A} 
 \eeq
 The minimum is clearly attained by  $q_A=q^*$, by  definition of $q^*$. 
 All in all, we have to choose the value of 
 $A$ such that $q_A=q^*$, the ``true'' saddle point.   Entropy and free energy give, for this value,  the correct results $S_G=S$ and $F_G=F$.
 We note that the  only purpose of the Gaussian ansatz at this stage is to fix the correct value of $q^*$ that dominates all  integrals, a purely geometric feature: indeed, any other ansatz (e.g. a delta function) would have given the correct result by a similar argument.
This is akin to the equivalence of different thermodynamic ensembles in infinite dimensional space.
 
If in the same example  we are interested in  the radial ``cage'' function $\r(r)$,
the result will be Gaussian if we use $\r_G$, while it will generically have tails $\sim e^{-dU}$ that are not Gaussian in the exact case. In other words, the
tails of the ``cage'' distribution are large deviation functions with (large) parameter $d$. They could also be computed in the Gaussian approximation by evaluating the free energy
as a function of the intensity $h$ of an additional potential $h x$: as usual large deviations control the thermodynamics in presence of an external field.
 
This paper is organized as follows: in section~\ref{sec:liquid} we write the replicated Van der Waals entropy in coordinates corresponding to the scalar products, as described above.
Sections~\ref{sec:J} and~\ref{sec:K} are devoted to the calculation of the Jacobians of these changes of coordinates, that will play the role of the term $r^{d-1}$ of polar coordinates in the example above.
(We need two Jacobians, corresponding to integrations in spaces of $m$ vectors $\{x_a\}$, and to $2m$ vectors $\{x_a,y_a\}$ -- as required by the Mayer function -- in terms of the corresponding scalar products). 
In section~\ref{sec:Mayer} we compute the Mayer function in terms of these coordinates. Thus, we obtain a complete expression (section~\ref{sec:Sq}) for the entropy, in terms of low-dimensional integrals, with
the dimension $d$ appearing as a parameter.
In sections~\ref{sec:Gaussian} and~\ref{sec:generic1RSB} we do the analogue of the previous paragraph: we compute the thermodynamic functions using the Gaussian ansatz and the exact solution taking saddle
points that become exact as $d \rightarrow \infty$. Both results coincide, thus validating the Gaussian ansatz.

\section{The replicated Van der Waals entropy}
\label{sec:liquid}

The starting point of our calculation is the free energy of a replicated liquid, where each atom
is replaced by a ``molecule'' made by one atom per each of the $m$ replicas~\cite{MP09,PZ10}. 
We denote by $\bar x = \{ x_1, \cdots, x_m\}$ the coordinate of such a molecule, each $x_a$ being
a vector in $d$-dimensional space. 
We assume that in the glass phase the molecule is well defined,
the typical distance between atoms in a molecule being of order $\sqrt{A}$ which is small at the glass
transition~\cite{PZ10}. Note that this is a non-trivial assumption: molecules might dissociate, especially close to the
glass transition, and lose their identity. However, we will show self-consistently in the end that this is not the case,
at least for $d\to\io$. Taking into account this effect in finite dimensions might be a non-trivial task.

The liquid state is described by a single copy of the system, $m=1$, with uniform density $\r$.
When $d\to\io$, its entropy (per particle) is given by Eq.~(\ref{1SS}) for $\r(x)=\r$,
which corresponds to the Van der Waals mean field equation:
\beq\label{Sliq}
s_{\rm liq} = \frac{\SS[\r]}N = 1 - \log \r + \frac{\r V_d D^d}2 = 1 - \log\r + 2^{d-1} \f \ .
\eeq
Here $V_d$ is the volume of a sphere of unit radius in $d$ dimensions, and $D$ is the sphere diameter,
and we introduced the {\it packing fraction} $\f = \r V_d (D/2)^d$.
It has been shown in~\cite{FRW85,FP99,PS00} that Eq.~\eqref{Sliq} is exact in the limit $d\to\io$, 
provided $2^d \f$ does not grow exponentially with $d$. If this is the case,
the other virial corrections are exponentially suppressed in $d$.

In the replicated liquid, atoms within a molecule can overlap, while atoms of different replicas belonging
to different molecules have the normal hard sphere interaction.
If $\sqrt{A} \ll D$, the molecule-molecule interaction is similar to the normal hard sphere interaction
and one can repeat the analysis of~\cite{FP99}.
The replicated liquid with integer $m \geq 1$ can thus be described in terms of a replicated
Van der Waals entropy given by Eq.~(\ref{virial})  (see ~\cite{PZ10,MK11}),
where $\r(\bar x)$ is the single molecule density, normalized to $V^{-1} \int d\bar x \r(\bar x) = \r$ where
$V$ is the system volume,
and $f(\bar x - \bar y)$ is the replicated Mayer function that describes the molecule-molecule interaction:
\beq
f(\bar x - \bar y) = -1 + \prod_{a=1}^m \theta(|x_a-y_a| - D) \ .
\eeq

We wish  to make use of the homogeneity of the molecular liquid, which implies that we
have to consider a generic translationally and rotationally invariant form of $\r(\bar x)$.
Let us start with translational invariance. We can perform a change of variables,
$X = m^{-1} \sum_a x_a$, $u_a = x_a - X$, where $X$ is the center of mass of the molecule and $u_a$
are the relative displacements with respect to $X$. 
Then
\beq
d\bar x = dX \ d\bar u \, m^d \, \d\left( \sum_{a=1}^m u_a \right) = dX \ \DD\bar u \ ,
\eeq
and translational invariance implies that $\r(\bar x)$ does not depend on $X$. We obtain
\beq\begin{split}
&\SS[\r(\bar u)] = V \int \DD\bar u \r(\bar u) [ 1 - \log \r(\bar u) ] + \frac{V}2 
\int \DD\bar u \DD\bar v \r(\bar u) \r(\bar v) \ol f(\bar u - \bar v) \ , \\
&\ol f(\bar u - \bar v) = \int dX f(X + \bar u - \bar v) \ ,
\end{split}\eeq
where $X + \bar u$ means adding $X$ to each component of $\bar u$.

Next we consider rotational invariance, which implies that $\r(\bar u)$ and $\ol f(\bar u)$
are functions of 
$q_{ab} = u_a \cdot u_b$ only. Let us define the following
quantities:
\beq\begin{split}
& q_{ab} = u_a \cdot u_b \\
& p_{ab} = v_a \cdot v_b \\
& r_{ab} = u_a  \cdot v_b \\
\end{split}\eeq
Note that translational invariance implies that $\sum_{a=1}^m q_{ab} = \sum_{b=1}^m q_{ab} = 0$ for all
row and columns.
Denoting $d\hat q = \prod_{a\leq q}^{1,m} dq_{ab}$, we have
\beq\begin{split}
\SS[\r(\hat q)]/V & = \int \DD\bar u \r(\hat q) [ 1 - \log \r(\hat q) ] + \frac{1}2 
\int \DD\bar u \DD\bar v \r(\hat q) \r(\hat p) \ol f(\hat q + \hat p - \hat r - \hat r^T) \\
& = \int d\hat q J(\hat q) \r(\hat q) [ 1 - \log \r(\hat q) ] + 
\frac{1}2 
\int d\hat q d\hat p d\hat r K(\hat q, \hat p, \hat r) \r(\hat q) \r(\hat p) \ol f(\hat q + \hat p - \hat r - \hat r^T) \ .
\end{split}\eeq
Here we introduced two Jacobians $J(\hat q)$ and $K(\hat q, \hat p, \hat r)$ that describe the change of variables from
$\bar u$ to $\hat q$, and from $(\bar u,\bar v)$ to $(\hat q, \hat p, \hat r)$ respectively. 
We compute them in the next sections. Before doing that, note that translational and rotational invariances imply that
the integrals are reduced from $\sim m \, d$ variables $\bar u$ to $\sim m (m-1)/2$ variables $\hat q$. This is crucial because
now the number of integration variables does not grow with $d$ and we can use saddle-point methods when $d\to\io$.

\section{The Jacobian $J$}
\label{sec:J}

\subsection{Definition}

The Jacobian $J(\hat q)$ is defined as:
\beq\label{Japp}
\begin{split}
J(\hat q) &= \int \DD\bar u \prod_{a\leq b}^{1,m} \d( q_{ab} - u_a \cdot u_b)
= m^d \int d\bar u \, \d\left( \sum_{a=1}^m u_a \right) \prod_{a\leq b}^{1,m} 
\d( q_{ab} - u_a \cdot u_b) \\
&= m^d \prod_{a=1}^m \d\left( \sum_{b=1}^m q_{ab} \right)
\int du_1 \cdots du_{m-1} \prod_{a\leq b}^{1,m-1} \d( q_{ab} - u_a \cdot u_b) \ ,
\end{split}\eeq
where the second line is obtained easily by manipulating the delta functions.

The delta functions take into account translational invariance. The last term instead
takes into account rotational invariance, and
it can be shown that
\beq\label{Japp2}
\int du_1 \cdots du_{m-1} \prod_{a\leq b}^{1,m-1} \d( q_{ab} - u_a \cdot u_b) = 
C_{m,d}  \ \ e^{\frac12 (d - m) \log \det \hat q^{m,m} } \ ,
\eeq
where $\hat q^{a,b}$ is the $(m-1)\times(m-1)$ matrix that is obtained by removing from $\hat q$
the $a$-th row and the $b$-th column.
In fact, Eq.~\eqref{Japp2} can be thought as a change of variables from a $d \times (m-1)$ matrix 
$\hat U = \{u_1, \cdots, u_{m-1}\}$
whose columns are the coordinates of the vectors $u_1 \cdots u_{m-1}$,
 to the $(m-1) \times (m-1)$ matrix 
$\hat q^{m,m} = \hat U^T \, \hat U$.
The corresponding Jacobian is given by Eq.~\eqref{Japp2}, see~\cite{MV09}.

Using this, we get the final result:
\beq\label{J}
\begin{split}
J(\hat q) &=  m^d \, C_{m,d} \prod_{a=1}^m \d\left( \sum_{b=1}^m q_{ab} \right)
\ \ e^{\frac12 (d - m) \log \det \hat q^{m,m} } \ .
\end{split}\eeq
Note that this form of $J$ is consistent with the fact that the
choice of the $m$-th row and column in (\ref{Japp}) is arbitrary: we could have chosen
any other row and column. But because the matrix $\hat q$ has the property that
the rows and columns add up to zero, it has the property that
$\det \hat q^{a,b} = \det \hat q^{m,m}$ for any $a,b$.

The normalization constant is
\beq\label{C}
\begin{split}
& C_{m,d} = 2^{1-m} \prod_{k=d-m+2}^d \Omega_k \ , \\
& \Omega_d = \frac{2 \pi^{d/2}}{\Gamma(d/2)} \ ,
\hskip1cm
\text{(the $d$-dimensional solid angle)}
\end{split}\eeq
as we show in next section.

\subsection{Calculation of the normalization constant}

Here we compute the normalization constant $C_{m,d}$. 
First we note that one can compute it directly for $m=2$ and $m=3$, using polar and bi-polar coordinates respectively:
\beq\begin{split}
& C_{2,d} = \frac{\Omega_d}2 \ , \\
& C_{3,d} = \frac{\Omega_d \Omega_{d-1}}{4} \ .
\end{split}\eeq
This already hints strongly at the form (\ref{C}).

Next we perform an asymptotic computation for large $d$ at fixed $m$.
For this, we write
(dropping for convenience the superscript $(m,m)$ on the matrix $\hat q$ that
here we consider here to be a generic $(m-1)\times(m-1)$ matrix):
\beq\begin{split}
(2 \pi)^{d(m-1)/2} & = 
\int du_1 \cdots du_{m-1} e^{-\frac12 \sum_{a=1}^{m-1} u_a \cdot u_a} \\
& = \int d\hat q \int du_1 \cdots du_{m-1} \  e^{-\frac12 \sum_{a=1}^{m-1} q_{aa}} 
\prod_{a\leq b}^{1,m-1} \d( q_{ab} - u_a \cdot u_b)\\
& = C_{m,d} \int d\hat q  \  e^{\frac12 (d - m) \log \det \hat q -\frac12 \sum_{a=1}^{m-1} q_{aa}}
\end{split}\eeq
The latter integral can be evaluated by a saddle point. 
Using $\frac{d}{dq_{ab}}  \log \det \hat q = (\hat q^{-1})_{ab}$,
the stationary equation reads:
\beq
- I + (d-m) \hat q^{-1} =0 \ , \hskip2cm
\hat q = (d-m) I \ .
\eeq
Next we expand
\beq
\hat q = (d-m) I + \hat t
\eeq
and we obtain, truncating the expansion in $\hat t$ at quadratic order:
\beq\begin{split}
(2 \pi)^{d(m-1)/2} & 
= C_{m,d} \int d\hat t  \  e^{\frac12 (d - m) 
\{ (m-1)\log(d-m) + \Tr[ \hat t/(d-m) - \hat t^2 / 2/ (d-m)^2 + \cdots    ]  \} 
-\frac12 \sum_{a=1}^{m-1} t_{aa} - \frac12 (d-m)(m-1)} \\
&= C_{m,d} (d-m)^{ (d - m) (m-1)/2}  e^{-\frac12 (d - m) (m-1)}
 \int d\hat t  \  e^{ - \frac1{4 (d-m)} \sum_{ab}^{1,m-1} t_{ab}^2 } \\
&= C_{m,d} (d-m)^{ (d - m) (m-1)/2}  e^{-\frac12 (d - m) (m-1)}
2^{(m-1)/2} \sqrt{2\pi (d-m)}^{m(m-1)/2}
\end{split}\eeq
We therefore get
\beq\label{Cas}
C_{m,d} = 2^{-\frac14(m-1)(2+m-2d)} e^{\frac12 (d-m)(m-1)} \pi^{-\frac14(m-1)(m-2d)}
(d-m)^{\frac14 m(m-1) - \frac12d (m-1)}
\eeq
It is easy to show that the limit for $d\to\io$ of this expression divided by Eq.~(\ref{C}) is given by 1.

\section{The Jacobian $K$}
\label{sec:K}

\subsection{Definition}

The Jacobian $K(\hat q,\hat p, \hat r)$ is defined as:
\beq\label{Kapp}
\begin{split}
K(\hat q, \hat p, \hat r) &= \int \DD\bar u \DD \bar v \prod_{a\leq b}^{1,m}  \d( q_{ab} - u_a \cdot u_b) \prod_{a\leq b}^{1,m} \d( p_{ab} - v_a \cdot v_b)
  \prod_{a, b}^{1,m} \d( r_{ab} - u_a \cdot v_b) \\
&= m^{2d} \int d\bar u d\bar v \, \d\left( \sum_{a=1}^m u_a \right) \d\left( \sum_{a=1}^m v_a \right) 
 \prod_{a\leq b}^{1,m}  \d( q_{ab} - u_a \cdot u_b) \prod_{a\leq b}^{1,m} \d( p_{ab} - v_a \cdot v_b)
  \prod_{a, b}^{1,m} \d( r_{ab} - u_a \cdot v_b) \\
&= m^{2d} \prod_{a=1}^m  \d\left( \sum_{b=1}^m q_{ab} \right)  \prod_{a=1}^m  \d\left( \sum_{b=1}^m p_{ab} \right)
 \d\left( r_{mm} - \sum_{a,b}^{1,m-1} r_{ab} \right)
 \prod_{a=1}^{m-1}  \d\left( \sum_{b=1}^m r_{ab} \right) \prod_{b=1}^{m-1}  \d\left( \sum_{a=1}^m r_{ab} \right) \\
& \times \int du_1 \cdots du_{m-1} dv_1 \cdots dv_{m-1}
 \prod_{a\leq b}^{1,m-1} \d( q_{ab} - u_a \cdot u_b)
 \prod_{a\leq b}^{1,m-1} \d( p_{ab} - v_a \cdot v_b)
  \prod_{a, b}^{1,m-1} \d( r_{ab} - u_a \cdot v_b)
\end{split}\eeq
where the last line is obtained easily by manipulating the delta functions.

We can define again a matrix $\hat U = \{ u_1, \cdots, u_{m-1}, v_1, \cdots, v_{m-1} \}$ of size $d \times 2(m-1)$
and a matrix $\hat Q = \hat U^T \, \hat U$ of size $2(m-1) \times 2(m-1)$, such that
\beq
\hat Q = \left[ 
\begin{matrix}
\hat q^{m,m} & \hat r^{m,m} \\
(\hat r^{m,m})^T & \hat p^{m,m}
\end{matrix}
\right]
\eeq
is obtained from the matrices $\hat q,\hat p, \hat r$ from which the $m$-th line and column has been removed.

Clearly we can write, using Eq.~(\ref{Japp2}) and calling $U_a$ the columns of the matrix $\hat U$:
\beq\begin{split}
& \int du_1 \cdots du_{m-1} dv_1 \cdots dv_{m-1}
 \prod_{a\leq b}^{1,m-1} \d( q_{ab} - u_a \cdot u_b)
 \prod_{a\leq b}^{1,m-1} \d( p_{ab} - v_a \cdot v_b)
  \prod_{a, b}^{1,m-1} \d( r_{ab} - u_a \cdot v_b)
= \\
& =  \int dU_1 \cdots dU_{2(m-1)} \prod_{a\leq b}^{1,2(m-1)} \d( Q_{ab} - U_a \cdot U_b ) = C_{2m-1,d} \ \ e^{\frac12 (d - 2m+1) \log \det \hat Q}
\end{split}\eeq

Using this, we get the final result:
\beq\label{K}
\begin{split}
K(\hat q, \hat p, \hat r) &=  m^{2d} \, C_{2m-1,d}  \ \ e^{\frac12 (d - 2m+1) \log \det \hat Q} \\
& \times
\prod_{a=1}^m  \d\left( \sum_{b=1}^m q_{ab} \right)  \prod_{a=1}^m  \d\left( \sum_{b=1}^m p_{ab} \right)
 \d\left( r_{mm} - \sum_{a,b}^{1,m-1} r_{ab} \right)
 \prod_{a=1}^{m-1}  \d\left( \sum_{b=1}^m r_{ab} \right) \prod_{b=1}^{m-1}  \d\left( \sum_{a=1}^m r_{ab} \right)
\end{split}\eeq

\section{The replicated Mayer function $\ol f$}
\label{sec:Mayer}

\subsection{General expression}

We now investigate the replicated Mayer function $\ol f(\bar u)$, which is defined as
\beq
\ol f(\bar u ) = \int dX \left\{  -1 + \prod_{a=1}^m \theta(|X + u_a | - D) \right\}
= - \int dX \,  \theta( D - \min_a | X + u_a | ) \ .
\eeq
The $u_a$ are $m$ vectors in $d$ dimensions. In the following we assume that $d>m$.
A remark that will be useful in the following is that when all $u_a=0$, $\ol f = - V_d D^d$, while
when the distance between each pair $|u_a - u_b | > D$, we have $\ol f = -m V_d D^d$.

We define $X_\parallel$ as the part of $X$ that lies in the hyperplane defined by the $u_a$ and $X_\perp$ the orthogonal part.
Then, recalling that $\Omega_d$ is the $d$-dimensional solid angle and $V_d=\Omega_d/d$,
\beq\begin{split}
\ol f(\bar u ) & =  - \int dX \,  \theta( D - \min_a | X + u_a | ) 
=  - \int dX \,  \theta( D^2 - \min_a | X + u_a |^2 ) \\
&= -\int d^{m}X_\parallel \ \ d^{d-m}X_\perp \ \ \theta( D^2 - \min_a\{ | X_\parallel + u_a |^2 + | X_\perp |^2 \} ) \\
&= - \Omega_{d-m} \int d^{m}X_\parallel \int_0^\io dx \, x^{d-m-1} \,  \theta( D^2 - x^2 - \min_a | X_\parallel + u_a |^2 ) \\
&= - \Omega_{d-m} \int d^{m}X_\parallel \int_0^{ \sqrt{D^2 - \min_a | X_\parallel + u_a |^2} }  dx \, x^{d-m-1} \\
& = - V_{d-m}  \int d^{m}X_\parallel \, (D^2  - \min_a | X_\parallel + u_a |^2 )^{(d-m)/2} 
\th(D^2  - \min_a | X_\parallel + u_a |^2 ) \\
& = - V_{d-m}  \int d^{m}X_\parallel \, \Theta_{d-m}(D^2  - \min_a | X_\parallel + u_a |^2 )
\end{split}\eeq
where we defined the function
\beq
\Th_{d-m}(x) = x^{(d-m)/2} \th(x) \ .
\eeq
While the above formula is always valid as long as $d>m$, 
for large $d$ the last integral is dominated by the points where $\min_a | X_\parallel + u_a | = 0$, which means that
$X_\parallel = - u_a$ for some $a$ (each value of $a$ defines a different saddle point). 
Observing that $\ol f = -V_d D^d$ when $u_a = 0 \ \forall a$, we can also write:
\beq\label{eq:fu}
\ol f(\bar u ) = - V_d D^d \, \frac{\int d^{m}X_\parallel \, \Theta_{d-m}(D^2  - \min_a | X_\parallel + u_a |^2 )}{\int d^{m}X_\parallel \, \Theta_{d-m}(D^2  - | X_\parallel |^2 )} \ .
\eeq

\subsection{Evaluation of $\ol f$ for $d\to\io$}

Let's consider first the case where the vectors $u_a$ are very large. In this case, 
if we write $X_\parallel = - u_a + \e$, for small $\e$ the minimum $\min_a | X_\parallel + u_a |$ 
will still be assumed in the same value of $a$ than
for $\e=0$, hence $\min_a | X_\parallel + u_a | = |\e|$.
Then we get
\beq\begin{split}
\ol f(\bar u) & \sim - V_{d-m} \sum_{a=1}^m \int d^{m}\e \, (D^2  - |\e|^2 )^{(d-m)/2} \th(D  - |\e| ) \\
& \sim - m \, V_{d-m} \Omega_m \int_0^D d\e \, \e^{m-1} (D^2-\e^2)^{(d-m)/2} \\
& \sim - m \, V_{d-m} \Omega_m D^d \frac{\G((d-m+2)/2) \G(m/2)}{d \G(d/2)} = - m V_d D^d
\end{split}\eeq
which implies that $\ol f(\bar u)$ is a constant exactly equal to minus the volume of $m$ hyperspheres, 
$-m \times V_d D^d$.
Note that the integral over $\e$ is dominated by a saddle-point at $\e \sim 1/\sqrt{d}$, as it can be
easily checked.
On the other hand, in the limit $|u_a|=0$ for all $a$, we trivially obtain
$\ol f(\bar u) = - V_d D^d$,
the volume of one hypersphere.

Therefore, the region where $\overline{f}$ has a non-trivial dependence on the $u_a$ is 
where the $u_a$ have a length proportional to $1/\sqrt{d}$.
We can define
$u_a = x_a D/\sqrt{d-m}$ and $X_\parallel = \ee D/\sqrt{d-m}$,
and we can write from Eq.~(\ref{eq:fu}):
\beq\label{flimit}
\begin{split}
\ol f(\bar u ) &  = - V_d D^d \ \frac{ \int d^{m}\ee \, \left(1  - \frac{ \min_a | \ee + x_a |^2 }{d-m} \right)^{(d-m)/2} \th\left(1  - \frac{ \min_a | \ee + x_a |^2 }{d-m} \right) }
{ \int d^{m}\ee \, \left(1  - \frac{ | \ee |^2 }{d-m} \right)^{(d-m)/2} \th\left(1  - \frac{  | \ee |^2 }{d-m} \right) }
\\
& \sim - V_{d} D^d \ \frac{ \int d^{m}\ee \, e^{  - \frac12 \min_a | \ee + x_a |^2  } }
{ \int d^{m}\ee \, e^{  - \frac12  | \ee |^2  } }
\end{split}\eeq
which is still of the order of $V_d D^d$ times a non-exponential factor that depends on the $u_a$.

We therefore conclude that $\ol f(\bar u) $ has the following scaling form when $d\to\io$:
\beq\label{olfscal}
\ol f(\bar u) = - V_d D^d \ \FF\left(\frac{\sqrt{d-m}}{D} \ \bar u \right) \ ,
\eeq
where
\beq\label{FF}
\FF(\bar x) =  \frac{ \int d^{m}\ee \, e^{  - \frac12 \min_a | \ee + x_a |^2  } }
{ \int d^{m}\ee \, e^{  - \frac12  | \ee |^2  } } =
 \int \frac{d^{m}\ee}{\sqrt{2 \pi}^m} \, e^{  - \frac12 \min_a | \ee + x_a |^2  }
\eeq
Note that when all $x_a=0$, $\FF=1$ as it should; and when each distance $|x_a - x_b|\to\io$, $\FF\to m$.

%

\section{Rotationally and translationally invariant expression of the replicated Van der Waals entropy}
\label{sec:Sq}

\subsection{Exact expression of the entropy}

Before proceeding, let us collect here the results obtained up to this point. 
We wrote the replicated Van der Waals entropy, taking into account explicitly rotational and translational invariance, as follows:
\beq\label{SSfinal}
\begin{split}
\SS[\r(\hat q)]/V = \int d\hat q J(\hat q) \r(\hat q) [ 1 - \log \r(\hat q) ] + 
\frac{1}2 
\int d\hat q d\hat p d\hat r K(\hat q, \hat p, \hat r) \r(\hat q) \r(\hat p) \ol f(\hat q + \hat p - \hat r - \hat r^T)
\end{split}\eeq
where
\beq\begin{split}
J(\hat q) &=  m^d \, C_{m,d} \prod_{a=1}^m \d\left( \sum_{b=1}^m q_{ab} \right)
\ \ e^{\frac12 (d - m) \log \det \hat q^{m,m} }
\end{split}\eeq
(here $\hat q^{m,m}$ is obtained by removing the $m$-th line and column from $\hat q$)
and
\beq
\begin{split}
K(\hat q, \hat p, \hat r) &=  m^{2d} C_{2m-1,d}  \ \ e^{\frac12 (d - 2m+1) \log \det \hat Q} \\
& \times
\prod_{a=1}^m  \d\left( \sum_{b=1}^m q_{ab} \right)  \prod_{a=1}^m  \d\left( \sum_{b=1}^m p_{ab} \right)
 \d\left( r_{mm} - \sum_{a,b}^{1,m-1} r_{ab} \right)
 \prod_{a=1}^{m-1}  \d\left( \sum_{b=1}^m r_{ab} \right) \prod_{b=1}^{m-1}  \d\left( \sum_{a=1}^m r_{ab} \right)
\end{split}\eeq
(here $\hat Q$ is obtained from the matrix $\left[ 
\begin{matrix}
\hat q & \hat r \\
\hat r^T & \hat p
\end{matrix}
\right]$ by removing the $m$-th and $2m$-th lines and columns)
and
\beq
\begin{split}
C_{m,d} = 2^{1-m} \prod_{k=d-m+2}^d \Omega_k 
\sim 2^{-\frac14(m-1)(2+m-2d)} e^{\frac12 (d-m)(m-1)} \pi^{-\frac14(m-1)(m-2d)}
(d-m)^{\frac14 m(m-1) - \frac12d (m-1)}
\end{split}\eeq

\subsection{Equation for $\r(\hat q)$}

Remember that $\r(\bar u)$ is normalized by $V^{-1} \int d\bar x \r(\bar x) = \int \DD \bar u \r(\bar u) = \r$.
Starting from Eq.~(\ref{SSfinal}) and differentiating with respect to $\r(\hat q)$, 
adding a Lagrange multiplier to ensure normalization,
we obtain the equation:
\beq
J(\hat q) \log \r(\hat q) =  \int d\hat r d\hat p K(\hat q, \hat p, \hat r)\r(\hat p) \ol f(\hat q + \hat p - \hat r - \hat r^T) +\l J(\hat q)
\eeq
Recall now that by definition:
\beq\begin{split}
& \int d\hat r d\hat p K(\hat q, \hat p, \hat r)\r(\hat p) = \int d\hat p d\hat r \int \DD\bar u \DD \bar v \prod_{a\leq b}^{1,m}  \d( q_{ab} - u_a \cdot u_b) \prod_{a\leq b}^{1,m} \d( p_{ab} - v_a \cdot v_b)
  \prod_{a, b}^{1,m} \d( r_{ab} - u_a \cdot v_b) \  \r(\bar v) = \\
  & =    \int \DD\bar u \DD \bar v \prod_{a\leq b}^{1,m}  \d( q_{ab} - u_a \cdot u_b)
 \  \r(\bar v) = \r \, J(\hat q) \ .
\end{split}\eeq
Therefore we can write the equation for $\r(\hat q)$ as follows:
\beq\label{eq:rho}
\log \r(\hat q) = \l + \r  \, \frac{\int d\hat r d\hat p K(\hat q, \hat p, \hat r)\r(\hat p) \ol f(\hat q + \hat p - \hat r - \hat r^T)}{\int d\hat r d\hat p K(\hat q, \hat p, \hat r)\r(\hat p)} \ ,
\eeq
where obviously the delta functions involving $\hat q$ in the expression of $K$ have to be formally simplified between numerator and denominator. The multiplier $\l$ is determined by the normalization condition.

\section{The Gaussian ansatz}
\label{sec:Gaussian}

Before moving to the general case, 
we show here that the computation above gives back exactly the results of~\cite{PZ10}
if a Gaussian ansatz is made for $\r(\hat q)$. Using this Gaussian ansatz, we evaluate
Eq.~(\ref{SSfinal}) using the saddle point method. In the next section we will show how this result can be obtained in fully generality.

We observe that because neither
$\log \r(\hat q)$ nor $\ol f(\hat q)$ are exponential in $d$, the saddle point is only 
determined by the Jacobians and by $\r(\hat q)$ in both terms of Eq.~(\ref{SSfinal}).
Therefore, $\hat r=\hat 0$ at the saddle point, and $\hat q = \hat p = \hat q^{sp}$, where
$\hat q^{sp}$ is the point where the exponential factor in $J(\hat q)\r(\hat q)$ is maximum.
Substituting this saddle point in Eq.~\eqref{SSfinal}, we obtain
\beq\label{Sgauss}
\begin{split}
\SS[\r(\hat q)]/N & \sim  1 - \log \r(\hat q^{sp}) + 
\frac{\r}2 
 \ol f(2 \hat q^{sp}) \\
& \sim  1 - \log \r(\hat q^{sp}) -
2^{d-1} \f \,
 \FF\left(\frac{d}{D^2}  2 \hat q^{sp} \right)
\end{split}\eeq
where the second line is obtained from the first by using Eq.~\eqref{olfscal}.

\subsection{The Gaussian form of $\r(\bar u)$ and the entropic term}

The Gaussian ansatz has the following form:
\beq
\r(\bar u) = \frac{\r m^{-d/2}}{(2\pi A)^{(m-1) d /2}} e^{-\frac{1}{2mA} \sum_{a<b}^{1,m} (u_a - u_b)^2} 
\hskip20pt
\Rightarrow
\hskip20pt
\r(\hat q) = \frac{\r m^{-d/2}}{(2\pi A)^{(m-1) d /2}} e^{-\frac{1}{2mA} \left( m \sum_{a=1}^m q_{aa} - \sum_{ab}^{1,m} q_{ab} \right) }  
\eeq
The first task is to compute the saddle point value of $\hat q$ that dominates all the integrals.
We have, using the delta functions contained in $J(\hat q)$ to manipulate the exponential term in $\r(\hat q)$:
\beq
\r =   \int d\hat q J(\hat q) \r(\hat q) \propto \int d\hat q  \prod_{a=1}^m \d\left( \sum_{b=1}^m q_{ab} \right)
\ \ e^{\frac12 (d - m) \log \det \hat q^{m,m} - \frac{1}{2A} \left( \sum_{a=1}^{m-1} q_{aa} + \sum_{ab}^{1,m-1} q_{ab} \right) } 
\eeq
At this point the integral over $q_{am}$ is eliminated by the delta functions and we are left with the $(m-1) \times (m-1)$ 
matrix $\hat q^{m,m}$.
The saddle point equation is, for $d\to\io$ and $a,b = 1, \cdots , m-1$:
\beq
( \hat q^{-1} )_{ab}^{sp} = \frac{1}{d \, A} (1 + \d_{ab} ) \ .
\eeq
The matrix is easily inverted and we obtain
\beq\label{qsp}
q_{ab}^{sp} = d \, A  \left( \d_{ab} -  \frac1m \right)
\eeq
It is easy to show using the conditions $\sum_{b=1}^m q_{ab} = 0$ imposed by the delta function that the formula above
holds for $a,b = 1, \cdots, m$. Indeed the saddle point values satisfy $q_{ab}^{sp} = \la u_a \cdot u_b \ra$ (where the average is over $\r(\bar u)$) so the same result could be
obtained from a direct computation.
We get
\beq
 1 - \log \r(\hat q^{sp})  = 1 - \log\r + \frac{d}{2} \log m + \frac{(m-1)d}{2} + \frac{(m-1)d}{2} \log(2 \pi A)
\eeq
which is the same result that can be obtained by an exact computation, the integrals being Gaussian in this case~\cite{PZ10}.

\subsection{The interaction term}
\label{sec:Gint}

Next, we  compute the term involving $ \ol f$ in the saddle point. 
Let us start with the following observation. Because $\ol f(\bar u)$ depends only on $\hat q$ thanks to rotational invariance,
all values of $\bar u$ that correspond to the same $\hat q$ give the same value of $\ol f(\bar u)$. This means that if we
want to compute $\ol f(\hat q)$, we can do that by choosing our favorite configuration of $\bar u$ that corresponds
to the chosen $\hat q$.

Therefore, for the saddle point (\ref{qsp}), we can choose any $u_a^{sp}$ that satisfy the conditions
\beq\begin{split}
&\sum_{a=1}^m u^{sp}_a = 0 \\
&u^{sp}_a \cdot u^{sp}_b = q_{ab}^{sp} = A \, d \left( \d_{ab} -  \frac1m \right)
\end{split}
\eeq
Remember that $u_a^{sp}$ are $d$-dimensional vectors.
A good choice is $(u_a^{sp})^b = \sqrt{A d} \left( \d_{ab} -  \frac1m \right)$ for their first $m$ components, $b=1,\cdots,m$, and zero 
for all the other components.

We therefore use this configurations of the $\bar u$ to compute $\ol f(\bar u)$.
We further define $\wh A$ by $A = D^2 \wh A/d^2$. 
Therefore, the
corresponding variables $\bar x$ that appear as the arguments of Eq.~(\ref{FF}) in
the saddle-point Eq.~(\ref{Sgauss}) have to satisfy
\beq
x^{sp}_a \cdot x^{sp}_b =\frac{d}{D^2} 2 q_{ab}^{sp} \sim 2 \wh A \left( \d_{ab} -  \frac1m \right) \ ,
\eeq
and therefore
\beq
(x_a^{sp})^b = \sqrt{2 \wh A} \left( \d_{ab} -  \frac1m \right) \ .
\eeq
With this choice, a short computation shows that
\beq
\min_a | \ee + x_a^{sp} |^2 = \sum_{b=1}^m (\ee^b)^2 - \frac{2 \sqrt{2 \wh A}}m \sum_{b=1}^m \ee^b
+2 \sqrt{2 \wh A} \min_a \ee^a + 2 \wh A \left(1 - \frac1m \right) \ .
\eeq
Therefore
\beq\begin{split}
\FF(\bar x^{sp}) & = 
 \int \frac{d^{m}\ee}{\sqrt{2 \pi}^m} \, e^{  - \frac12 \min_a | \ee + x^{sp}_a |^2  } \\
& = m \, e^{- \wh A \frac{m-1}m } \, \int_{-\io}^\io 
\frac{d\ee}{\sqrt{2 \pi}} e^{-\frac12 \ee^2 -\sqrt{2 \wh A} \, \ee \, \frac{m-1}m }
\left( \int_{\ee}^\io \frac{d\h}{\sqrt{2 \pi}}  e^{-\frac12 \h^2 + \frac{\sqrt{2 \wh A}}m \h} \right)^{m-1} \\
& = m \, e^{- \wh A \left(\frac{m-1}m\right)^2 } \, \int_{-\io}^\io 
\frac{d\ee}{\sqrt{2 \pi}} e^{-\frac12 \ee^2 -\sqrt{2 \wh A} \, \ee \, \frac{m-1}m }
\left[ \frac12 \left( 1 + \erf \left( \frac{\sqrt{2 \wh A} - \ee m}{\sqrt{2} m}  \right) \right) \right]^{m-1} \\
& = m  \int_{-\io}^\io \frac{d\ee}{\sqrt{2 \pi}} e^{-\frac12 \ee^2 }
\left[ \frac12 \left( 1 + \erf\left( \frac{ \sqrt{2 \wh A} - \ee }{\sqrt{2} }  \right) \right) \right]^{m-1}  \ .
\end{split}\eeq
It is useful to define
\beq
\GG_m(\wh A) = 1 - \FF(\bar x^{sp})  
= 1 - m  \int_{-\io}^\io \frac{d\ee}{\sqrt{2 \pi}} e^{-\frac12 \ee^2 }
\left[ \frac12 \left( 1 + \erf\left( \frac{ \sqrt{2 \wh A} - \ee }{\sqrt{2} }  \right) \right) \right]^{m-1}
 \ .
\eeq

\subsection{The Gaussian result}

The final result of the Gaussian computation is therefore, at the leading order for $d\to\io$:
\beq\label{GaussianS}
\SS[\r(\hat q)]/N = 
1 - \log\r + \frac{d}{2} \log m + \frac{(m-1)d}{2} + \frac{(m-1)d}{2} \log\left(\frac{2 \pi D^2 \wh A}{d^2}\right) 
- 2^{d-1} \f [1 - \GG_m(\wh A) ] \ .
\eeq
It coincides exactly with the result of~\cite{PZ10}; the explicit expression of $\GG_m(\wh A)$ that
was given in~\cite{PZ10} is different, but it is exactly equivalent to the present one
(actually the present one is much easier to compute numerically).

\section{The generic 1-step replica symmetry broken saddle point}
\label{sec:generic1RSB}

Finally we analyze the structure of the generic solution for $\r(\hat q)$ in the large $d$ limit without assuming a Gaussian form for the density function.
Like in the Gaussian case, we want to
evaluate the integrals in Eq.~(\ref{SSfinal}) and Eq.~(\ref{eq:rho})
via a saddle point. We will show that we will recover the results coming from the Gaussian case.

\subsection{Structure of the saddle point}

First we have to derive the saddle point equations.
Let us suppose that
\beq\begin{split}
& \r(\hat q) = e^{-\Omega(\hat q)} \\
& \omega(\hat q^{m,m}) = \Omega\left(q_{ab}, \  q_{am} = -\sum_{b=1}^{m-1} q_{ab}, \ q_{mb} = -\sum_{a=1}^{m-1} q_{ab}, \
q_{mm} = \sum_{a,b}^{1,m-1} q_{ab} \right) \ .
\end{split}\eeq
The 1-step replica symmetric breaking (1RSB) solution consists in assuming that
$\Omega(\hat q)$ has a replica symmetric (RS) structure. Indeed, this corresponds to 1RSB because
the present real replica scheme describes what happens inside one of the 1RSB blocks~\cite{Mo95,MP09,PZ10}.

First we want to determine the saddle-point value of $\hat q$, that dominates the normalization of $\r(\hat q)$. We have
\beq
\r =   \int d\hat q J(\hat q) \r(\hat q) \propto \int d\hat q  \prod_{a=1}^m \d\left( \sum_{b=1}^m q_{ab} \right)
\ \ e^{\frac12 (d - m) \log \det \hat q^{m,m} -   \omega(  \hat q^{m,m} ) } \ .
\eeq
Therefore the delta functions allow to eliminate the $m$-th line of $q_{ab}$, and the 
saddle point equations for the remaining variables $q_{ab}$, with $a,b=1\cdots m-1$, are determined by the maximization
of the exponential factor for $d\to\io$:
\beq
\frac{d}2 ( \hat q^{-1} )_{ab}^{sp} = \frac{d  \omega(  \hat q ) }{d q_{ab}} 
= \frac{d\Omega}{dq_{ab}} + \frac{d\Omega}{dq_{mm}} - \frac{d\Omega}{dq_{am}} - \frac{d\Omega}{dq_{mb}}
\ .
\eeq
Because $\Omega(\hat q)$ has a RS structure, we have
$ \frac{d\Omega}{dq_{ab}}  = \Omega_0(\hat q) + \frac{d^2}{2\Omega_1(\hat q)}  \d_{ab}$, and
\beq\label{pqr_sp}
\begin{split}
& ( \hat q^{-1} )_{ab}^{sp} = \frac{d}{ \Omega_1(\hat q^{sp})} (1 + \d_{ab}) \ , \\
& q_{ab}^{sp} = \frac{\Omega_1(\hat q^{sp})}{d} \left( \d_{ab} -  \frac1m \right) \ . 
\end{split}\eeq

Consider now the integral:
\beq
\r^2 = \int d\hat q d\hat p d\hat r K(\hat q, \hat p, \hat r) \r(\hat q) \r(\hat p)  \ .
\eeq
A very similar procedure leads to the following saddle point equation:
\beq\label{Qsp}
\frac{d}2 \hat Q^{-1} \equiv \frac{d}2
\left[ 
\begin{matrix}
\hat q^{m,m} & \hat r^{m,m} \\
(\hat r^{m,m})^T & \hat p^{m,m}
\end{matrix}
\right]^{-1}
=
\left[ 
\begin{matrix}
\frac{d^2}{2 \, \Omega_1(\hat q)} (1 + \d_{ab})  & \hat 0 \\
\hat 0 & \frac{d^2}{2 \, \Omega_1(\hat p)} (1 + \d_{ab}) 
\end{matrix}
\right]
\eeq
whose solution is $r_{ab}=0$ and $q_{ab} = p_{ab} = q^{sp}_{ab}$, both equal to the solution
of Eq.~(\ref{pqr_sp}).
Finally, we are interested in the integral entering in Eq.~(\ref{eq:rho}), that is the same as the last one
but without the condition on $\hat q$. One obtains the same as Eq.~(\ref{Qsp}) but without the upper left
block of the matrix. However, the equation for $\hat r$ is still solved by $\hat r=\hat 0$, then the equation on
$\hat p$ decouples from $\hat q$ and leads to the same solution as in Eq.~(\ref{pqr_sp}).

We conclude that the integrals in Eq.~(\ref{SSfinal}) and Eq.~(\ref{eq:rho}) can be evaluated via a saddle point
and the result is
\beq\label{SSspfinal}
\begin{split}
\SS[\r(\hat q)]/N & \sim  1 - \log \r(\hat q^{sp}) + 
\frac{\r}2 
 \ol f(2 \hat q^{sp}) \\
& \sim  1 - \log \r(\hat q^{sp}) -
2^{d-1} \f \,
 \FF\left(\frac{d}{D^2}  2 \hat q^{sp} \right)
\end{split}\eeq
and
\beq\label{eq:rhosp}
\begin{split}
\log \r(\hat q) & = \l + \r \ol f(\hat q + \hat q^{sp}) \\ &= \l - 2^d \f  \,
 \FF\left(\frac{d}{D^2} (\hat q +  \hat q^{sp} )\right)
 \ .
\end{split}\eeq
Therefore, to conclude the calculation we have to determine $\hat q^{sp}$ and $\l$.

Before proceeding, two remarks are in order. First of all, when the distance between atoms in a molecule is large,
the $q_{ab}$ are large, and $\FF \to m$ as discussed in Section~\ref{sec:Mayer}. Hence in this limit 
$\r(\hat q) \sim \exp(-2^d \f m)$, and because $2^d \f \propto d$ at the glass transition~\cite{PZ10}, we see that $\r(\hat q)$ goes
to a very small constant that vanishes exponentially with $d$. Hence, in the $d\to\io$ limit the molecules are well 
defined in the glass phase, while for finite $d$ there is an exponentially small probability of dissociation. This guarantees
that the molecular liquid is a good description of the glass phase for large $d$.
The second remark is that the choice of a given RSB ansatz is self-consistent. We assumed at the beginning a RS structure for
$\Omega(\hat q)$; then we obtained that $\hat q^{sp}$ has a RS structure as given in Eq.~(\ref{pqr_sp}); and finally that, self-consistently,
$\Omega(\hat q) = - \log \r(\hat q)$ has a RS structure as given by Eq.~(\ref{eq:rhosp}). We could consider a 1RSB structure for $\Omega(\hat q)$
(corresponding to a 2RSB computation in the real replica scheme) and we would have obtained self-consistently the same structure for $\hat q^{sp}$.
Saddle points characterized by many steps of RSB could be needed to describe the metastable states of lower density~\cite{MR04}.

\subsection{Saddle point equation}

We now have to solve the saddle point equation Eq.~(\ref{pqr_sp}).
We note that, defining $\wh A^{sp} = \Omega_1(\hat q^{sp}) / D^2$, we can rewrite 
Eq.~(\ref{pqr_sp}) as a closed equation for the scalar parameter $\wh A^{sp}$:
\beq\label{eq:A_sp}
\begin{split}
&q_{ab}^{sp} = \frac{D^2 \wh A^{sp}}{d} \left( \d_{ab} -  \frac1m \right) \ , \\
&\wh A^{sp} = \frac1{D^2} \Omega_1\left[  \frac{D^2 \, \wh A^{sp}}{d} \left( \d_{ab} -  \frac1m \right)  \right] \ ,
\end{split}\eeq
Furthermore, if we define a function
\beq
h(\wh A) = \Omega\left[  \frac{D^2 \, \wh A}{d} \left( \d_{ab} -  \frac1m \right)  \right] \ ,
\eeq
then we have
\beq
\frac{d h}{d\wh A}(\wh A^{sp}) = \sum_{ab}^{1,m} \frac{d \Omega}{d q_{ab}}(\hat q^{sp}) \, \frac{D^2}d  \left( \d_{ab} -  \frac1m \right) 
=  \sum_{ab}^{1,m} \left(  \Omega_0(\hat q^{sp}) + \frac{d^2}{2\Omega_1(\hat q^{sp})}  \d_{ab}   \right) 
\frac{D^2}d  \left( \d_{ab} -  \frac1m \right) 
= \frac{d \, D^2 \, (m-1)}{2 \Omega_1(\hat q^{sp})} \ ,
\eeq
therefore
\beq
\Omega_1(\hat q^{sp}) =  \frac{d \, D^2 \, (m-1)}{2 h'(\wh A^{sp})}
\eeq
and the equation for $\wh A^{sp}$ becomes
\beq
\wh A^{sp} = \frac{d \, (m-1)}{2 h'(\wh A^{sp})} \ .
\eeq

The function $h(\wh A)$ can be computed by using Eq.~(\ref{eq:rhosp}), that gives
$\Omega(\hat q) = - \l + 2^d \f  \,
 \FF\left(\frac{d}{D^2} (\hat q +  \hat q^{sp} )\right)$. Then
\beq
h(\wh A) = - \l + 2^d \f  \,
 \FF\left[ (\wh A + \wh A^{sp})  \left( \d_{ab} -  \frac1m \right)    \right] \ .
\eeq
The computation of this function can be done exactly as we did in the Gaussian case,
in section~\ref{sec:Gint}, and leads to the following result:
\beq
 \FF\left[ (\wh A + \wh A^{sp})  \left( \d_{ab} -  \frac1m \right)    \right] = 
1- \GG_m[ (\wh A + \wh A^{sp})/2 ] \ .
\eeq
Therefore
\beq
h'(\wh A^{sp}) = -2^{d-1} \f \GG_m'(\wh A^{sp}) \ ,
\eeq
and finally the equation for $\wh A^{sp}$ is
\beq
\wh A^{sp} = -\frac{d \, (m-1)}{2^{d} \f \, \GG_m'(\wh A^{sp}) } \ .
\eeq
It is easy to check that this equation is exactly the same that is obtained by maximization of
the Gaussian free entropy Eq.~(\ref{GaussianS}).
Therefore, the generic saddle point equation coincides with the Gaussian one.

\subsection{The computation of $\l$}

The last ingredient that we need to compute the replicated free entropy in the generic case is
the value of $\l$. Indeed, combining Eq.~(\ref{SSspfinal}) and Eq.~(\ref{eq:rhosp}), and recalling
that $ \FF\left(\frac{d}{D^2}  2 \hat q^{sp} \right) = 1 - \GG_m(\wh A^{sp})$,
 we obtain
\beq\label{appl}
\begin{split}
\SS[\r(\hat q)]/N 
& =   1 - \l +
2^{d-1} \f \, [1 - \GG_m(\wh A^{sp})] \ .
\end{split}\eeq

The factor $\l$ has to be computed by imposing the normalization of $\r(\hat q)$:
\beq
\r =  \int d\hat q J(\hat q) \r(\hat q)
= e^{\l} \  m^d \, C_{m,d}
\int d\hat q
 \prod_{a=1}^m \d\left( \sum_{b=1}^m q_{ab} \right)
\ \ e^{\frac12 (d - m) \log \det \hat q^{m,m} - 2^d \f  \,
 \FF\left(\frac{d}{D^2} (\hat q +  \hat q^{sp} )\right) } \ .
\eeq
It is enough to evaluate the integral at the saddle point level to get the part of $\l$ that is proportional
to $d$. Corrections to the saddle-point give corrections to $\l$ that are at most proportional to $\log d$ and
will be neglected here.
We get from Eq.~(\ref{eq:A_sp})
\beq
\log \det (\hat q^{sp})^{m,m} = (m-1) \log(D^2 \wh A^{sp}/d) - \log m \ .
\eeq
Recalling that from Eq.~(\ref{Cas}) we have, neglecting corrections proportional to $\log d$:
\beq
\log C_{m,d} = \frac{d}2 (m-1) \log(2 \pi e) - \frac{d}2 (m-1) \log d \ ,
\eeq
and once again that $ \FF\left(\frac{d}{D^2}  2 \hat q^{sp} \right) = 1 - \GG_m(\wh A^{sp})$,
we obtain the equation for $\l$:
\beq\begin{split}
\log \r & = \l + d \log m + \frac{d}2 (m-1) \log(2 \pi e) - \frac{d}2 (m-1) \log d + \frac{d}2 \big[ (m-1) \log(D^2 \wh A^{sp}/d) - \log m \big] 
- 2^d \f \big[ 1 - \GG_m(\wh A^{sp}) \big]  \\
& = \l + \frac{d}2 \log m + \frac{d}2 (m-1) + \frac{d}2 (m-1) \log(2 \pi D^2 \wh A^{sp}/d^2)  
- 2^d \f \big[ 1 - \GG_m(\wh A^{sp}) \big]  \ ,
\end{split}\eeq
where $\log d$ terms are neglected.
Plugging this in Eq.~(\ref{appl}) we finally obtain
\beq
\SS[\r(\hat q)]/N 
=   1 - \log \r
 + \frac{d}2 \log m + \frac{d}2 (m-1) + \frac{d}2 (m-1) \log(2 \pi D^2 \wh A^{sp}/d^2)  
- 2^{d-1} \f \, [1 - \GG_m(\wh A^{sp})] \ ,
\eeq
which coincides exactly with the Gaussian result, Eq.~(\ref{GaussianS}).

\section{Conclusions}

We have derived the large dimensional limit for the statistical mechanics of dense amorphous hard spheres.  We have shown that a so-called Gaussian ansatz gives 
the correct result for the thermodynamic functions. The results previously obtained with such an ansatz are  thus validated, at least near the transition where a 
one-step replica symmetry breaking scheme is expected to suffice. 
We notice that the same computation would also apply to the Bethe lattice model of~\cite{MKK08} in the high coordination and large dimension limit.

There is however reason to suspect that the actual exact result has an infinite number of breakings, at least in the limit of high pressure: 
first of all,  the generic situation with systems which have a transition to a one-step solution is to have a further transition to a phase with more -- eventually infinite  
levels of replica symmetry breaking. More physically, we know that hard spheres at large pressure develop many soft vibrational modes \cite{wyart,LNSW10,He10} due to isostaticity.
This is true not only of the equilibrium states, but also of the metastable ``J-point'' states. Now, 1RSB equilibrium states have a spectrum with no soft modes -- and this is true of 
all but the very highest metastable states. A full replica symmetry breaking scheme would naturally bring in soft modes, as happens for example in the case of spin-glasses.
Perhaps the transition into such a phase  also brings in isostaticity at high pressure, something the 1RSB solution displays only for the equilibrium states~\cite{PZ10}.
The study of the stability of the 1RSB solution is under investigation.
 
As mentioned in the simple example of the introduction, the large $d$ calculation we have presented cannot be expected to yield the exact result for the cage distribution,
especially its tails. A more detailed calculation, always within this framework and based on Eq.~\eqref{eq:rhosp}, is possible for the tails of exponentially small probability, both at the glass
transition and at jamming. Hopefully such calculation will be able to reproduce the numerical results of~\cite{CIPZ12}, where a large non-Gaussian tail has been
detected in the self part of the van Hove function, which coincides with the cage distribution. It was found that this tail is not reduced on increasing dimension and seems to
persist even for $d\to\io$, suggesting that it could be described by mean field theory.
This will be the subject of a future paper. 

To conclude, let us mention that it would be nice to reproduce the results obtained here without using replicas,
i.e. by finding directly the amorphous solutions of Eq.~\eqref{1SS}. This approach (which is also called Density Functional Theory or DFT)
has been pursued in~\cite{KW87}, under the assumptions that {\it (i)} the density field $\r(x)$ is the sum of Gaussians centered around
amorphous reference positions $R_i$ and {\it (ii)} the structure factor of the $R_i$ is the same as those of the liquid. The results of~\cite{KW87} are close but
not exactly equivalent to the ones of replica theory. It is likely that hypothesis {\it (i)} is not needed thanks to the same mechanism that was exposed in
the introductions and is at work in the replica calculation. However, it is less easy to refrain from using hypothesis {\it (ii)} in DFT, and it is likely that
this hypothesis is false. In fact, the liquid structure becomes akin to the one of the ideal gas when $d\to\io$, while it is likely that the $R_i$ remain
correlated, at least for neighboring particles (as it is clear for instance in the jamming limit of infinite pressure).
The replica method is indeed designed to integrate over the unknown $R_i$ to get rid of them and avoid hypothesis {\it (ii)}.
Reconciling DFT with the replica method requires a way to solve for the $R_i$, which has yet to be found.

\bibliographystyle{mioaps}
\bibliography{HS}

\end{document}